# GenAIT: An Objective Generative AI Literacy Test for High School Students and Its Associations with LLM Use and Perceptions

**Brett Puppart[1,2*], Kristjan-Julius Laak[1], Jaan Aru[1]**

[1]Institute of Computer Science, University of Tartu, Estonia

[2]Institute of Psychology, University of Tartu, Estonia

**Corresponding author:** brett.puppart@ut.ee

**Abstract**

There is growing international interest in generative AI (GenAI) literacy and its assessment among high school students, but objective assessment in this population remains limited. Also, little is known about how demonstrated knowledge about generative technologies relates to students' perceptions and use of LLMs. This article reports on the development and validation of the GenAI Literacy Test (GenAIT), an 18-item multiple-choice test measuring high school students' conceptual knowledge about GenAI, with content spanning technical, practical, and human-impact domains. We also examined its associations with perceived usefulness, perceived ease of use, and general and school-related LLM use. Expert review of relevance, clarity, and comprehensiveness provided evidence of content validity. In a large-scale survey of 7,432 Estonian high school students, we evaluated the psychometric functioning of GenAIT using confirmatory factor analysis, classical test theory, and item response theory. Results supported approximate unidimensionality and good fit of a three-parameter logistic model (RMSEA = .013, TLI = .987, CFI = .990, SRMSR = .021). Score reliability was broadly adequate (marginal reliability = .72, KR-20 = .69), although precision varied substantially across the latent trait. GenAIT is therefore more suitable for group-level research than high-stakes individual classification. We found no significant association between GenAIT scores and perceived ease of use, a very small negative association with perceived usefulness (Spearman's $\rho$ = −.05), and a small to medium negative association with both general (Spearman's $\rho$ = −.16) and school-related LLM use frequency (Spearman's $\rho$ = −.19). These findings highlight the need to distinguish AI adoption and favorable perceptions from demonstrated conceptual understanding of generative technologies.

# 1 Introduction

Generative artificial intelligence (GenAI) is a technology with the potential to significantly disrupt how people live, work, and learn (Feuerriegel et al., 2024). Citizens should therefore be educated to adapt to a world increasingly shaped by these technologies. Just as the spread of digital technologies prompted the development of digital literacy education (Buckingham, 2006), there is now a growing need for AI education (Miao et al., 2024; OECD & European Union, 2026). To inform these educational efforts, objective assessment is needed to measure what students know about generative technologies.

However, the current literature on AI literacy assessment for K-12 students has important limitations. First, most of the existing tests measure general AI literacy, with few focusing specifically on GenAI, despite the prominent role of generative technologies in current AI use (e.g., Carolus et al., 2023; Chiu et al., 2024b). Second, most GenAI literacy tests rely on subjective self-report assessment, which may be good for measuring perceived competence, but simultaneously lead to inaccuracies when we are interested in real knowledge and understanding. Finally, the existing objective measures have not been developed and validated on high school students (Jin et al., 2025; Markus et al., 2025; Rismanchian et al., 2026). This marks an important limitation, as large language model (LLM) adoption among these students remains high (Granström & Oppi, 2025), and large-scale initiatives, including the Estonian AI Leap, specifically target the high school population (AI Leap Foundation, 2025; Aru & Laak, 2025).

The lack of objective instruments also limits our understanding of how GenAI literacy relates to other constructs, including perceived usefulness, perceived ease of use, and LLM use frequency − variables central to the technology acceptance model (TAM; Davis, 1989). In educational contexts, understanding the link between GenAI literacy and LLM use is important, because the patterns of LLM use can directly impact learning outcomes (Bastani et al., 2025). Perceptions of LLMs are also relevant, as they have been associated with ChatGPT use (Duong et al., 2023) and intention to learn about AI (Gu et al., 2026). Nevertheless, little is known about these relationships when GenAI literacy is assessed objectively among high school students.

The present study therefore addresses two related aims. First, we develop and collect initial validity evidence of an objective test measuring the conceptual knowledge component of GenAI literacy among high school students. Second, we use this measure to provide initial evidence on how demonstrated GenAI knowledge relates to perceptions of LLMs and the frequency of general and school-related LLM use.

# 2 Literature review

## 2.1 Defining GenAI literacy

Literacy was originally understood in relation to the skills and knowledge required to read, understand, and use written language (UNESCO, 2004). As societies became more digitalized, researchers proposed the construct of digital literacy to describe skills and knowledge related to digital tools (Buckingham, 2006). More recently, advances in AI have made AI literacy an important construct to guide research and policymaking. Long and Magerko (2020) defined AI literacy as a set of competencies that enables individuals to critically evaluate AI technologies, collaborate with AI, and use AI as a tool in different areas of life.

Since then, numerous definitions and frameworks have been proposed, reflecting the lack of universal consensus on the boundaries of AI literacy (OECD & European Union, 2026). One key area of friction is whether AI literacy should be understood primarily in cognitive terms, centered on knowledge and skills (e.g., in Chiu et al., 2024a), or whether it should, more broadly, also include affective and enacted components, such as attitudes, critical reflection, and the application of AI-related knowledge and skills in practice (e.g., in Miao et al., 2024; OECD & European Union, 2026). Chiu et al. (2024a), for example, distinguished AI literacy from the broader construct of AI competency. In their framework, literacy is primarily related to the knowledge and skills required to use, understand, and critically evaluate AI, whereas competency is also about confidence, reflective capacity, and the ability to apply this knowledge in practice.

AI is an umbrella term encompassing substantially different technologies, including predictive and generative AI, which differ in how they function, are used, and may fail (Narayanan & Kapoor, 2024). Therefore, AI literacy as a construct may be too broad when the specific type of AI under consideration is left undefined. To distinguish competencies specifically related to generative technologies from the broader AI literacy, the term "GenAI literacy" has been used (Annapureddy et al., 2025; Bozkurt, 2024; Zhang & Magerko, 2025). GenAI literacy can be understood as a focused extension of AI literacy, rather than a completely distinct construct. Its distinctiveness is partly linked to the central role of natural-language prompting to support effective collaboration (Liu et al., 2023). It also involves risks that are particularly salient in generative contexts, including inappropriate reliance on AI-generated outputs (Klingbeil et al., 2024) and the rapid production and spread of misinformation (Zhou et al., 2023).

To identify the scope of conceptual knowledge relevant to GenAI literacy in high school students, we relied on the tri-partite framework draft provided by the Council of Europe (2026). It includes technological (How Artificial Intelligence works and how it might be developed), practical (How Artificial Intelligence can be used effectively), and human (The impact of AI on humans, human rights, democracy and the rule of law) dimensions. Gu and Ericson (2025) reached a similar model of AI literacy in the age of GenAI through a synthesis of 124 AI literacy studies. These domains guided content coverage and were not intended to define separate subscales. GenAIT operationalizes the conceptual knowledge component of GenAI literacy, sampling knowledge across technical, practical, and human domains. In Figure 1, we present a summary of the conceptualization and scope of GenAIT.

**The technical domain** contains the basics about how GenAI works and solves problems. A student should acquire a basic understanding of what goes on "under the hood" of LLMs. Including, but not limited to, understanding that GenAI systems are technologies that produce novel data (Feuerriegel et al., 2024), that token prediction is the fundamental mechanism of how LLMs work (Vaswani et al., 2017), and recognizing that increasing the size of training data and model parameters improves performance (Brown et al., 2020; Kaplan et al., 2020). This technological knowledge should illuminate the differences between human thinking and LLM output generation, which is important because linguistic fluency in LLM outputs does not reflect human-like understanding or lived experience (Aru et al., 2023; Mitchell, 2023).

**The practical domain** is related to students' understanding of their active role in getting GenAI to do what they want, including via prompting and critical evaluation of its outputs. Output quality can depend substantially on the prompt (Liu et al., 2023), which requires sufficient context and specificity. Consequently, errors such as ambiguity, biased language, and not accounting for potential interference of previous messages may lead to unsatisfactory results (Giray, 2023; Gupta et al., 2024). At the same time, some problems with LLM outputs are difficult or impossible to overcome with efficient prompting. LLM outputs are associated with frequent problems such as sycophancy, biases, and hallucinations (Borji, 2023; Sharma et al., 2024). Students should therefore be aware of these problems and understand the importance of fact-checking and critical thinking during GenAI use.

**The human domain** focuses on the potential implications of GenAI on human cognition, society, and the environment. Over-reliance is a frequent pattern in human-AI interaction (Klingbeil et al., 2024), which might negatively impact our critical thinking, creativity, and learning (Aru, 2025; Bastani et al., 2025; Kosmyna et al., 2025; Zhai et al., 2024). On the societal level, GenAI poses risks to democratic processes (Summerfield et al., 2025), online information ecosystems (Zhou et al., 2023; Quelle & Bovet, 2024), algorithmic echo chambers (Kirk et al., 2024), and social inequality through a widening digital divide (Wang et al., 2025a). Additionally, GenAI is not climate neutral, as both model training and inference consume substantial amounts of energy and water and produce carbon emissions (Li et al., 2025; Luccioni et al., 2024; O'Donnell & Crownhart, 2025).

**Figure 1**

*Framework of GenAI literacy and scope of GenAIT*

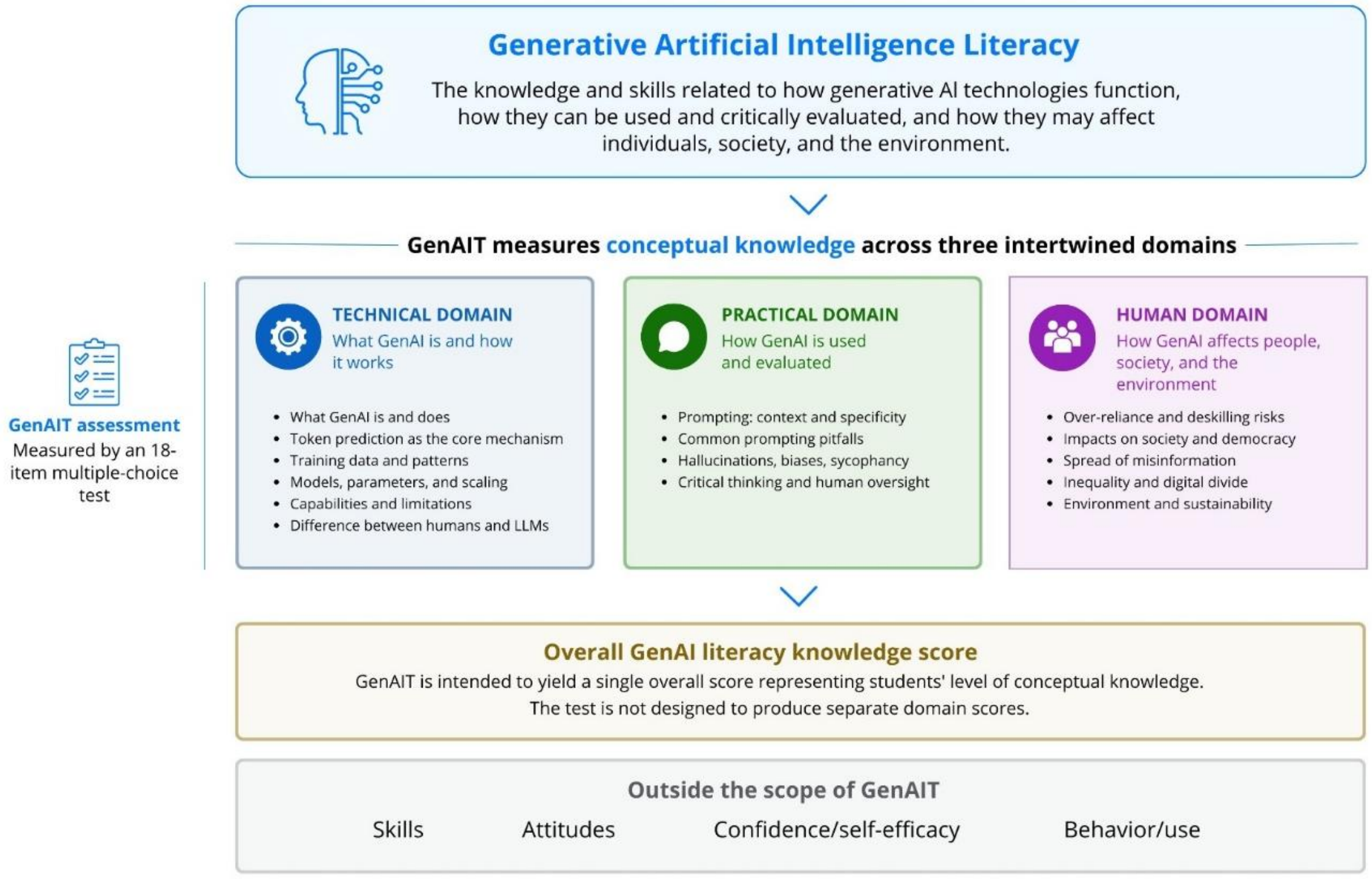

## 2.2 Measuring GenAI literacy

A variety of validated instruments exist for measuring AI literacy in different contexts and target populations (Lintner, 2024). Similarly, several instruments exist that measure GenAI literacy, competence, or knowledge exclusively or include it as a separate subdimension (Durak et al., 2026; Gökçearslan et al., 2024; Gümüş & Kara, 2025; Jin et al., 2025; Lee & Park, 2024; Lee et al., 2026; Markus et al., 2025; Noh et al., 2024; Wang et al., 2025b; Wu & Yang, 2025; Yan et al., 2026).

Collectively, these tests cover important themes and target populations. However, important limitations remain in the assessment of GenAI literacy, especially among the K-12 student population. As a general problem that is also prevalent in general AI literacy assessment (Lintner, 2024), most GenAI literacy tests are self-report questionnaires. While subjective measures may provide value in assessing aspects of the broader conceptualization of AI literacy that includes affective and enacted components, they may introduce bias when assessing real knowledge and skills. Research on metacognitive calibration has shown that people may inaccurately assess their own knowledge or competence (Kruger & Dunning, 1999), while self-report measures are also susceptible to broader response biases (Paulhus & Vazire, 2007).

These concerns are supported by emerging empirical evidence indicating a mismatch between subjective and objective measures of AI literacy. Jin et al. (2025) found that their objective GenAI literacy test outperformed self-report in predicting real-life AI-assisted task performance. Bewersdorff et al. (2025b) found in a systematic review that subjective measures yield systematically stronger correlations with affective variables than objective measures. Zhang et al. (2026) examined this issue directly by creating self-report and objective measures of AI literacy on matching dimensions and found only weak associations between them. Together, these findings suggest that objective assessment provides important information about AI and GenAI literacy that may not be adequately captured by self-report measures.

Among the few available instruments that allow objective assessment of GenAI-related knowledge or literacy, none appear to have been specifically developed and validated on high school students. The Generative AI Literacy Assessment Test (GLAT; Jin et al., 2025) is a 20-item multiple-choice measure designed for higher education students and structured around Ng et al.'s (2021) four dimensions of AI literacy: know and understand, use and apply, evaluate and create, and ethics. The AI Competency Objective Scale (AICOS; Markus et al., 2025) is a broader 51-item measure of AI literacy that includes GenAI literacy along with other AI competencies. While it is intended for use across different populations, GenAI represents only one component of the broader construct, potentially limiting its ability to comprehensively assess the GenAI-related knowledge and skills that may be particularly relevant for high school students. Together, these developments represent important progress in objective GenAI literacy assessment, but a methodological gap remains: a validated, developmentally appropriate measure of high school students' GenAI literacy is still needed.

Finally, GenAI literacy should ideally be assessed in the students' native language to avoid confounds via foreign language proficiency. However, none of the existing GenAI literacy instruments, both objective and subjective, have been validated in the Estonian language. This is an important gap, as the national AI Leap initiative aims to foster AI literacy among Estonian high school students (AI Leap Foundation, 2025). The present study therefore addresses both broader and local needs in GenAI literacy assessment: more broadly, by developing an objective GenAI literacy test specifically for high school students, and locally, by providing an Estonian-language instrument for this population.

### 2.3 GenAI literacy, LLM perceptions, and use

The technology acceptance model (TAM) proposes that perceived ease of use and usefulness shape attitudes and behavioral intentions, ultimately influencing technology adoption (Davis, 1989). The framework has been applied to understand the adoption of many technologies, including computer-based information systems (Davis, 1989), smartphones (Park & Chen, 2007), and, more recently, AI tools (Schiavo et al., 2024). While positive perceptions of ChatGPT have been linked to greater use (Duong et al., 2023), it is less clear how GenAI literacy, or what individuals know and understand, fits into the picture.

The existing evidence on these associations is inconclusive. Self-report measures of AI literacy have generally demonstrated positive associations with TAM-related variables. People who report higher AI literacy also report higher perceived usefulness (Al-Abdullatif & Alsubaie, 2024; Faiz et al., 2026; Ho & Van Le, 2026; Schiavo et al., 2024), ease of use (Obadă et al., 2026; Schiavo et al., 2024), and use these technologies more frequently (Bewersdorff et al., 2025b). Evidence from objective measures is more mixed. Studies on broader AI literacy and frequency of AI use have shown no significant association (Bewersdorff et al., 2025a) or a negative association (Tully et al., 2025). By contrast, a recent study focusing exclusively on GenAI literacy found a positive association with frequency of LLM use among undergraduate students (Rismanchian et al., 2026). To our knowledge, objective measures have not been used to investigate the association of AI literacy with perceived usefulness or perceived ease of use.

These relationships may also depend on the context of AI use and the target population. Existing studies have focused primarily on university students, and it is unclear to what degree these results generalize to younger students. Beyond general adoption of LLMs, an important question for educational research is how much students use these tools for learning-related purposes, as it is possible these two forms of use may diverge. A student may, for example, use LLMs frequently in everyday life while using them more selectively for schoolwork. To our knowledge, evidence on the association between objectively measured AI or GenAI literacy and school-related LLM use frequency among high school students is currently lacking. In the present study, we examine these associations using an objective measure of GenAI literacy.

## 3 The present study

In this article, we report on the development and validation of the Generative Artificial Intelligence Literacy Test (GenAIT), an objective measure of high school students' conceptual knowledge about GenAI. Validation proceeded through multiple stages, including expert review, a pilot study, and a large-scale study of Estonian high school students. The final instrument consists of 18 multiple-choice items covering technical, practical, and human content domains. It is intended for group-level research, including the examination of correlates of this knowledge and describing aggregate response patterns. Both summative raw scores and latent trait scores derived from a 3PL IRT model can be used for interpretation. The psychometric research questions were:

**RQ1**: To what extent do expert judgments support the relevance, clarity, and comprehensiveness of the initial GenAIT item pool for the defined content domain?

**RQ2**: What are the internal structure, item functioning, and score reliability and precision of the final version of GenAIT?

Beyond instrument development and validation, our aim was to examine how objectively measured GenAI literacy is associated with variables related to the TAM. The corresponding research question was:

**RQ3**: How is GenAI literacy, as measured by GenAIT, associated with perceived usefulness of LLMs, perceived ease of LLM use, and the frequency of general and school-related LLM use?

## 4 Materials and methods

We developed GenAIT informed by the standard guidelines for educational and psychological testing (AERA et al., 2014). Our research method involved multiple stages, which are outlined in Figure 2. Primary evidence from test content was obtained from an expert panel. Preliminary evidence from relation to external variables came from the pilot study. Internal structure and score reliability were assessed based on the data from the research project uncovering the outcomes related to the AI Leap initiative.

**Figure 2**

*Development and validation process of GenAIT*

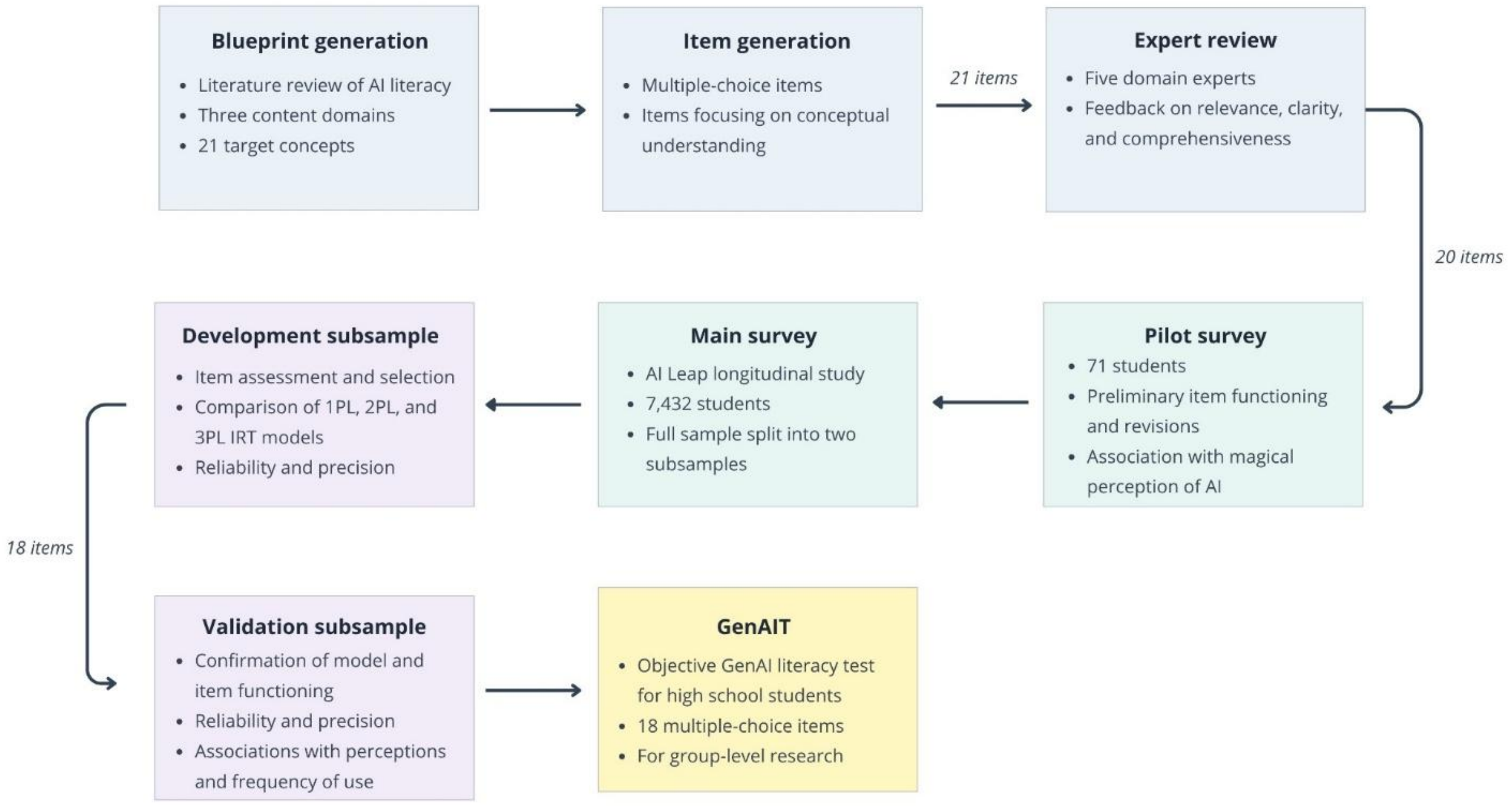

*Note.* Numbers on the arrows indicate the number of items retained at each stage.

## 4.1 Content blueprint and item generation

Prior to instrument development, the construct should be clearly defined, and the content domain specified (AERA et al., 2014). GenAIT was intended to measure the conceptual knowledge component of GenAI literacy, consisting of three content domains: technical, practical, and human. Importantly, these domains were specified to ensure representative coverage of GenAI literacy and were not intended to be separate latent dimensions or subscale scores.

Following an extensive literature review, we identified candidate concepts for each content domain. Through discussions between the authors of this paper, concepts that were judged as relevant and age-appropriate were included. Priority was given to concepts that were expected to remain educationally relevant despite rapid technological advances, rather than current model-specific capabilities. This process resulted in an initial pool of 21 target concepts, 7 in each content domain. A short introduction to the content domains, including concepts, and sample items is presented in Table 1.

**Table 1**

*GenAIT content domains, concepts, and sample items*

| Domain | Target concepts | Sample item |
|---|---|---|
| Technical | Recognizing machine learning, **Recognizing GenAI**, LLM training data, Scaling laws, Token prediction, Training-data regularities, Limits of LLM understanding | An artificial intelligence model has been trained on millions of paintings and can identify with high accuracy whether a painting is Impressionist or Expressionist in style. Why is the system described not a generative AI system? (Item 2)<br>A. It does not analyze text<br>B. It can solve only one narrow task<br>**C. It cannot create new paintings**<br>D. It has been trained on too little data |
| Practical | Sycophancy, Assessing output credibility, Few-shot prompting, Context interference, Prompt specificity, Prompt-induced bias, **Prompt-dependent capabilities** | A person complains that large language models only write very general and neutral essays. What is the person most likely doing wrong? (Item 14)<br>A. They are writing prompts that are too complicated<br>**B. They are writing prompts that are too vague**<br>C. Their prompts contain contradictory information<br>D. They do not understand that language models find it difficult to express clear positions and arguments |
| Human | Energy demands, Water use, **Misinformation**, Digital divide, Information accessibility, Cognitive offloading, Algorithmic echo chambers | What distinguishes generative AI from earlier technologies in relation to misinformation? (Item 17)<br>A. Generative AI reduces the amount of misinformation online by enabling automated fact-checking<br>B. Generative AI mainly affects text-based misinformation and has little impact on images or videos<br>C. AI-generated misinformation is easier to detect because it follows predictable patterns<br>**D. Generative AI makes it possible to create large amounts of convincing misinformation without specialized skills or significant time investment** |

*Note.* All items and target concepts of GenAIT are presented in Supplementary material S1. Sample items have been translated from Estonian to English. In the process of item selection, three target concepts were removed from the final version of GenAIT: LLM training data, few-shot prompting,

and information accessibility. Bold text in target concepts corresponds to the sample item. The correct answer option for each sample item is marked in bold.

Items were generated through an iterative cycle of drafting, review, and refinement conducted primarily by five members of the Artificial and Natural Intelligence Lab at the University of Tartu. Each target concept was operationalized by a multiple-choice item in the Estonian language that had four answer options with one correct response (see Supplementary material S1 for items and answer key). The items were designed to measure conceptual understanding. Accordingly, most items reflected scenarios requiring respondents to apply conceptual knowledge to realistic GenAI situations rather than recall isolated facts. Distractors were intentionally designed to reflect misconceptions or partially correct intuitions and were based on the researchers' understanding of the current literature.

The work was conducted in a shared online document, where each researcher could independently provide items and leave constructive comments about the draft. Disagreements about item correctness, distractor quality, and wording were discussed until consensus was reached.

### 4.2 Expert review

Content validity is the degree to which test items are relevant and sufficient for measuring the underlying construct. This includes evaluation of test items by a panel of experts or judges from the target population (Boateng et al., 2018). We relied on subject-matter experts to assess content representation and item clarity; students' interpretations of the items were not directly examined at this stage.

The review was conducted in January 2026. Experts were recruited via email and provided with a link to a questionnaire on the LimeSurvey web platform. Eligibility criteria included holding at least a master's degree, having multiple years of professional experience in their respective field, and demonstrating expertise in the field of AI and/or AI education. In total, we contacted eight individuals, of whom five agreed to participate, which is consistent with common recommendations for content validation studies (Lynn, 1986). The final expert panel consisted of three AI or data science experts, one educational technologist, and one psychologist.

The questionnaire and items were presented in Estonian. Upon opening the questionnaire, experts were first informed about our definition of GenAI literacy, the intended purpose of the instrument, and the target population. Both qualitative and quantitative feedback were collected to evaluate the content validity of GenAIT. Quantitative feedback included assessing the relevance and clarity of each item on a four-point forced-choice scale. For item relevance, answer options were coded as 1 = not relevant, 2 = somewhat relevant, 3 = quite relevant, 4 = highly relevant; for item clarity, as 1 = not clear, 2 = somewhat clear, 3 = quite clear, 4 = very clear. After rating individual items, they rated the comprehensiveness of the test by indicating agreement with the statement "I believe that this test covers the important concepts of high school students' generative AI literacy", where 1 = strongly disagree, 2 = somewhat disagree, 3 = somewhat agree, 4 = strongly agree. Experts were also invited to provide qualitative feedback on individual items and the test, as well as provide any important concepts that were missing in their opinion.

Item revision was guided by both qualitative and quantitative feedback. The primary metric of content validity was the item-level content validity index (I-CVI), which indicates the proportion of experts rating the relevance of an item as 3 or 4. With five raters, values of 1.00 are generally considered necessary for establishing excellent item-level content validity (Polit & Beck, 2006).

Analogously to the I-CVI, we also calculated item clarity indices. Items with I-CVI or clarity index values below 1.00 were flagged for review. In total, eight items were flagged for review. Seven of these went through minor revision to the item stem or individual distractors to remove factual ambiguities or problems with clarity. One item (Item 19) had both low reported relevance and clarity, and it could not be revised without substantial alterations in meaning. Therefore, this item was excluded from the instrument. See Supplementary material S3.2 for more detailed results of the expert review.

The I-CVI ratings were averaged across all items to a single scale level content validity index (S-CVI-Ave), with values above .90 indicating excellent content validity (Polit & Beck, 2006). For the initial 21-item version of GenAIT, we obtained an S-CVI-Ave value of .96 and a similarly computed clarity index of .89. Following the main validation study, we retrospectively recalculated these metrics for the 18 retained items of GenAIT and found an S-CVI-Ave of .99 and a clarity index of .90. These retrospective indices summarize the original expert ratings of the retained items; the revised wording of the final instrument was not independently re-rated.

Comprehensiveness ratings were generally favorable. One expert did not fill out this part of the questionnaire, resulting in a panel size of four. Experts generally agreed that the test covered important concepts of high school students' GenAI literacy (Mean = 3.25 on a four-point agreement scale). In the qualitative feedback, two experts commented on missing concepts. These included broader categories of statistical modeling and ethics, with no specific concepts identified.

## 4.3 Pilot study

In the pilot study, conducted in February 2026 before the main validation study began, we assessed feasibility, preliminary item functioning, and the expected association between GenAIT scores and magical perception of AI. Students from two schools were invited to participate; none participated in the main validation study. Out of 77 respondents, 71 remained after predefined exclusions. Participants completed the 20-item version of GenAIT (following the removal of one item after expert review) and a translated seven-item measure of magical perception of AI (developed by Tully et al., 2025). Preliminary item functioning was assessed using CTT indices, including item difficulty, discrimination, and item-total correlations. Item discrimination was evaluated by comparing the upper and lower 27% of respondents, following Kelley (1939) and Kumar et al. (2021).

GenAIT showed adequate internal consistency (KR-20 = .77, 95% CI [.69, .84]). Item 15 demonstrated poor discrimination and a low item-total correlation. Distractor analysis identified a dysfunctional distractor, which we revised before the main study. As predicted, we found a negative correlation between the 20-item version of GenAIT and magical perception of AI (Spearman's $\rho = -.35$, 95% CI [−.54, −.13], $p = .003$, $n = 71$). Following the main validation study, we retrospectively recalculated this association with the retained 18 items in the pilot sample and found a similar association (Spearman's $\rho = -.30$, 95% CI [−.50, −.07], $p = .010$, $n = 71$). A summary of the pilot study is provided in Supplementary material S3.3.

## 4.4 Main validation study

### 4.4.1 Participants

This study was part of the first round of a research project linked to AI Leap. All Estonian high schools were invited to participate in the study. Students in Grades 10 and 11 from 100 participating schools completed the LimeSurvey web questionnaire. The total number of respondents was 8,698.

After applying the pre-defined exclusion criteria (see 4.4.4 Data preparation and statistical analyses), 1,266 (~14.6%) were removed, leaving a final analytic sample of 7,432 students. The analytic sample comprised approximately 36.6% of all students enrolled in Grades 10 and 11 in Estonia in the 2025/2026 school year (Statistikaamet, 2026). Student demographic data, including age and gender, and school identifiers were not available to researchers due to data privacy considerations.

### 4.4.2 Materials

The main study administered the 20-item post-pilot form of GenAIT. The full AI Leap impact questionnaire contained a wide range of validated scales and tasks from learning motivation to creativity and critical thinking. In the present work, we analyzed only some of these that were relevant to our current research questions. All the measures that we analyzed here (except for GenAIT) were self-report items or scales. These include frequency of LLM use for schoolwork and in general, perceived usefulness of LLMs, and perceived ease of LLM use.

General and school-related LLM use frequency were each measured with one item on a scale from 1 (I do not use AI in general, for anything/for schoolwork) to 7 (Many times a day). Perceived usefulness of LLMs and perceived ease of LLM use are subscales adapted from the AI TAM scale developed by Yilmaz et al. (2023). Both subscales include three items that were translated into Estonian by a professional translator. The response format was converted into a six-point Likert-type scale, where 1 = Strongly disagree and 6 = Strongly agree. The scores of these items were summed to make two composite indices, ranging from 3 to 18: perceived usefulness of LLMs and perceived ease of LLM use. Supplementary material S1 contains GenAIT items and answer key. Supplementary material S2 presents all the other questionnaires used in this study.

### 4.4.3 Procedure

The study period lasted from February to the beginning of April 2026. Each school completed the questionnaire at their freely chosen times within this period in a computer class. Students were instructed to complete the survey on their own with no external aid. Upon opening the questionnaire on the LimeSurvey web platform, students read the study information and provided their informed consent. The questionnaire was composed of three parts and took around 30 minutes to complete. All of the materials used in these analyses were presented in the last part, as they were all thematically AI-related. Answering questions was not required to move forward.

Students initially completed the items of the previously described external measures and then the GenAIT. To minimize ordering effects, items within multi-item scales were presented in a randomized order. The order of GenAIT answer options was randomized, except for Item 3 that had a logical order of responses. To detect inattentive respondents, we introduced two instructed response attention checks after the 7th and 14th items in the randomized GenAIT item sequence, and screened completion times of GenAIT items. Mean completion time of 20 GenAIT items in the final analytic sample was 9.28 minutes, median = 9.19, SD = 4.07.

### 4.4.4 Data preparation and statistical analyses

All statistical analyses were conducted in R (R Core Team, 2025), using base R and the tidyverse (Wickham et al., 2019), psych (Revelle, 2025), lavaan (Rosseel et al., 2025), mirt (Chalmers, 2012), and boot (Canty & Ripley, 2025) packages.

Respondents missing a response to one or more of the 20 candidate GenAIT items were excluded from the analysis, resulting in the removal of 295 respondents. We then removed respondents demonstrating insufficient engagement based on both attention-check questions and completion times, as combining multiple indicators can improve the identification of inattentive responders (Abbey & Meloy, 2017). Participants were removed if they (a) failed both attention checks or (b) failed one attention check and had a natural-log-transformed GenAIT completion time more than two standard deviations below the mean log completion time. This cutoff was calculated after excluding incomplete GenAIT responses but before engagement exclusions and corresponded to approximately 1.79 minutes. In total, 761 respondents failed both attention checks and 210 failed one attention check and completed the test too quickly, resulting in the exclusion of another 971 respondents, leaving a final analytic sample of 7,432. We examined sensitivity to the exclusion criteria by comparing descriptives, IRT model fit, score reliability, and associations with external variables under the original, permissive, and strict rules (Supplementary material S3.1.5).

The remaining respondents were randomly split into approximately equal development and validation subsamples. The development subsample ($n$ = 3,737) was used for item selection and comparison of competing IRT models. The validation subsample ($n$ = 3,695) was held out from these decisions and used to independently examine the internal structure, reliability, item functioning, and model fit of the final instrument.

IRT was used as the primary framework for evaluating the psychometric quality of GenAIT. IRT models the probability of a correct response as a nonlinear function of an underlying latent trait and provides item-level estimates (Rusch et al., 2017). Three logistic IRT models of increasing complexity are commonly used for dichotomous items. The 1PL model allows the difficulty ($b$) of items to vary but assumes equal discrimination, 2PL adds the discrimination parameter ($a$) allowing item slopes to differ, and a 3PL adds the pseudo-guessing parameter ($g$) by raising the lower asymptote (Baker & Kim, 2017).

Unidimensional IRT models considered here assume a single underlying trait (Bichi & Talib, 2018), which we tested with a one-factor categorical confirmatory factor analysis (CFA) model with weighted least squares mean and variance adjusted (WLSMV) estimation. Model fit was considered acceptable when RMSEA was below .06, CFI and TLI were above .95, and SRMR was below .08 (Hu & Bentler, 1999). We evaluated the assumption of local independence (Bichi & Talib, 2018) with Yen's Q3 statistic (Yen, 1984). Values below .20 were considered non-problematic (Chen & Thissen, 1997). We evaluated the assumption of monotonicity (Täht et al., 2025) using empirical item plots showing the proportion of correct responses across increasing rest-score levels, with each item excluded from its own score. Differential item functioning could not be evaluated because we lacked demographic grouping variables.

Item functioning was assessed in both subsamples by examining the standardized factor loadings, IRT parameters, and item-fit statistics. Loadings ≥ .30 were interpreted as indicating at least a moderate association between the item and a factor (Tavakol & Wetzel, 2020). Discrimination was interpreted as: 0 = none, 0.01–0.34 = very low, 0.35–0.64 = low, 0.65–1.34 = moderate, 1.35–1.69 = high, ≥ 1.70 = very high (Baker & Kim, 2017). Item-fit was assessed with Orlando–Thissen $S\text{-}X^2$ statistic, where significant $p$-values signal potential misfit (Orlando & Thissen, 2000). To adjust for multiple testing, we used the Benjamini-Hochberg false discovery rate (FDR) correction. Because chi-square statistics are sensitive in large samples, item-fit was judged in conjunction with item-level RMSEA, parameter estimates, factor loadings, local dependence, and content relevance. No item was removed based on a single criterion.

1PL (discrimination fixed at 1), 2PL, and 3PL models were compared only in the development subsample. For this, we compared nested models using the likelihood-ratio test where statistically significant differences signal improved fit. We also looked at AIC and BIC, which balance goodness-of-fit with model complexity. Lower values are considered superior. The final item set and IRT model were fixed before examining the results of the validation subsample.

Absolute IRT model fit was assessed in both subsamples using the M2 statistic and associated approximate-fit indices. Because the M2 test is sensitive to large sample sizes, greater emphasis was put on approximate-fit indices (CFI, TLI, RMSEA, and SRMSR), item-fit statistics, and local dependence diagnostics rather than on statistical significance alone. RMSEA, CFI, and TLI were interpreted based on the criteria described above, and SRMSR was interpreted analogously to SRMR. In the validation subsample, assumption checks, model fit, and item functioning were repeated independently using the prespecified final instrument and model. Respondents' latent trait estimates were obtained from the final 3PL model using expected a posteriori (EAP) scoring.

Reliability of the final GenAIT instrument was evaluated using KR-20 and IRT marginal reliability. Cronbach's alpha was used for the multi-item Likert-type external measures. As a descriptive benchmark, values around .70 were considered adequate for group-level research (Bland & Altman, 1997). Measurement precision across levels of the latent trait was examined using the IRT test information curve. Conditional reliability was calculated from test information ($I(\theta)$) based on the formula: $I(\theta)/(I(\theta) + 1)$ (Nicewander, 2018).

Correlation analyses were conducted in the validation subsample to examine associations among GenAIT scores, perceived usefulness, perceived ease of use, and general and school-related LLM use frequency. We calculated a Spearman correlation matrix because LLM use frequency was measured using ordered response categories and the relationships of interest were expected to be monotonic rather than necessarily linear. Correlations were obtained using pairwise complete observations. Pointwise 95% percentile bootstrap confidence intervals were estimated using 5,000 respondent-level resamples. Statistical significance was evaluated at $p < .05$ after adjusting $p$-values across the ten pairwise correlations using the FDR procedure. Following Funder and Ozer (2019), correlation magnitudes were interpreted as approximately .05 = very small, .10 = small, .20 = medium, .30 = large, and .40 = very large.

Raw GenAIT total scores were used for these analyses because they are easier to implement and interpret. To assess whether this scoring choice affected the findings, we calculated Spearman's correlation between raw total scores and EAP scores to examine similarity in students' rank ordering. We also compared their respective correlations with the four external variables. In a post hoc sensitivity analysis, we excluded Item 12, refitted the 3PL model, and compared model fit, reliability, and correlations based on the revised raw scores with the original results. The results of both sensitivity analyses are provided in Supplementary material S3.1.4.

### 4.4.5 Ethics

This study was part of a project which was approved by the Research Ethics Committee of the Estonian Research Council (nr TEET207, dated 11th of February 2026). Participation was not mandatory. All participants read detailed instructions and provided written informed consent before participation. The results were pseudonymized and all demographic information removed by the Estonian Ministry of Education. The cleaned data was uploaded to a secure server of the University of Tartu for the data analysis.

### 4.4.6 Use of generative AI tools

During manuscript preparation, the authors used ChatGPT (OpenAI) as a conversational aid to discuss ideas and code, clarify methodological and theoretical concepts, assist with translation, and improve the language and readability of the manuscript. The tool was not used to generate or analyze study data or to make substantive research decisions. All AI-assisted content was critically reviewed and edited by the authors, who verified its accuracy and take full responsibility for the final manuscript.

# 5 Results

## 5.1 Item selection and model comparison

Item selection was conducted in the development subsample ($n$ = 3,737), using the 20-item version of GenAIT. A one-factor CFA using WLSMV estimation supported the approximate unidimensionality of the 20-item version (RMSEA = .023, TLI = .957, CFI = .962, SRMR = .045). A 3PL model provided the best fit among the candidate IRT models (Table 2), and local independence was supported (maximum Yen's Q3 = .045).

Following item analysis, Items 3 and 10 were removed. Item 3 displayed a negative standardized CFA loading, negative IRT discrimination, significant S-$X^2$ statistic, and lack of monotonicity. Although Item 10 showed strong IRT discrimination, it had a weak standardized CFA loading and was somewhat conceptually redundant with other prompting-related items, motivating its removal. Four other items were flagged by predefined criteria: Items 9 and 20 had factor loadings below .30, Item 8 had a discrimination parameter below .65, and Item 12 had a significant S-$X^2$ statistic and a downturn in the empirical proportion of correct responses at higher rest scores, suggesting a possible departure from monotonicity. These items were reviewed and retained after considering the extent of the deviation and content validity. See Supplementary material S3.1.1 for all item parameters and fit indices.

Model comparisons were repeated for the resulting 18-item version in the development subsample. Comparison of 1PL, 2PL, and 3PL models indicated that the 3PL model fit significantly better than the simpler models and produced lower AIC and BIC values (Table 2). Overall, the model demonstrated acceptable fit (RMSEA = .016, TLI = .982, CFI = .986, SRMSR = .020), although the M2 test was statistically significant (M2(117) = 226.383, $p$ < .001). A one-factor CFA again supported approximate unidimensionality (RMSEA = .023, TLI = .962, CFI = .967, SRMR = .044), and local independence remained supported (maximum Yen's Q3 = .045). At the item level, the only newly flagged item was Item 14, which had a very minor deviation from the predefined discrimination threshold. See Supplementary material S3.1.1 for all item parameters and fit indices.

**Table 2**

*Comparison of IRT models for the initial 20-item and final 18-item versions of GenAIT in the development subsample (n = 3,737)*

| Model | Items | AIC | BIC | logLik | $\Delta\chi^2$(df) | $p$ |
|---|---|---|---|---|---|---|
| 1PL | 20 | 89256.35 | 89380.87 | −44608.17 | | |
| 2PL | 20 | 88370.34 | 88619.38 | −44145.17 | 926.01 (20) | <.001 |
| 3PL | 20 | 88003.37 | 88376.93 | −43941.69 | 406.96 (20) | <.001 |
| 1PL | 18 | 81509.03 | 81621.10 | −40736.51 | | |
| 2PL | 18 | 81052.98 | 81277.11 | −40490.49 | 492.05 (18) | <.001 |
| 3PL | 18 | 80730.16 | 81066.37 | −40311.08 | 358.81 (18) | <.001 |

*Note.* AIC = Akaike information criterion; BIC = Bayesian information criterion; logLik = log likelihood. Model comparisons were conducted separately within each item set. Likelihood-ratio tests compare each model with the preceding model within the same item set. The discrimination parameter was fixed at 1 in the 1PL models.

## 5.2 Validation

A one-factor CFA with WLSMV estimation supported the approximate unidimensionality of GenAIT in the validation subsample ($n$ = 3,695). All model fit indices met the recommended thresholds: RMSEA = .025, TLI = .958, CFI = .963, SRMR = .047. Again, the 3PL model demonstrated acceptable fit (RMSEA = .013, TLI = .987, CFI = .990, SRMSR = .021), while the M2 test was statistically significant (M2(117) = 193.267, $p < .001$). The largest off-diagonal Q3 value was .035; therefore, problematic local dependence was not detected. Empirical rest-score plots showed broadly increasing response proportions for most items, with greater uncertainty at extreme scores, while Item 12 showed a downturn at higher rest scores (Supplementary material S3.1.3). All discrimination parameters were positive, ranging from 0.59 to 4.23 (Table 3). Item difficulties ranged from −1.71 to 2.05. Pseudo-guessing parameters ranged from 0.00 to 0.33 (item characteristic curves for each item are presented in Supplementary material S3.1.2). From the items flagged in the development subsample, three continued to meet at least one flagging criterion: Items 8, 9, and 12. After adjusting for multiple testing, Item 1 now also demonstrated a significant S-$X^2$ statistic, but the misfit was relatively minor. See Supplementary material S3.1.1 for all item parameters and fit indices.

**Table 3**

*Item parameters of the 3PL IRT model in the validation subsample*

| Item | Domain | % correct | *a* [95% CI] | *b* [95% CI] | *g* [95% CI] |
|---|---|---|---|---|---|
| Item 1 | Technical | 32.1 | 1.33 [0.90, 1.75] | 1.40 [1.24, 1.57] | 0.16 [0.10, 0.21] |
| Item 2 | Technical | 39.7 | 1.21 [0.88, 1.55] | 0.97 [0.78, 1.17] | 0.16 [0.08, 0.23] |
| Item 4 | Technical | 36.9 | 1.96 [1.37, 2.55] | 1.32 [1.19, 1.45] | 0.25 [0.21, 0.28] |
| Item 5 | Technical | 22.1 | 4.23 [2.53, 5.93] | 1.47 [1.38, 1.57] | 0.15 [0.13, 0.16] |
| Item 6 | Technical | 19.9 | 2.97 [2.19, 3.75] | 1.60 [1.49, 1.71] | 0.13 [0.11, 0.14] |
| Item 7 | Technical | 48.1 | 1.65 [1.17, 2.14] | 0.79 [0.61, 0.96] | 0.27 [0.20, 0.33] |
| Item 8 | Practical | 45.6 | 0.59 [0.49, 0.69] | 0.34 [0.05, 0.64] | 0.00 [−0.07, 0.08] |
| Item 9 | Practical | 21.4 | 2.17 [1.33, 3.01] | 2.05 [1.84, 2.27] | 0.17 [0.15, 0.19] |
| Item 11 | Practical | 57.3 | 0.76 [0.54, 0.98] | −0.37 [−1.19, 0.45] | 0.03 [−0.25,0.30] |
| Item 12 | Practical | 81.8 | 1.06 [0.92, 1.20] | −1.71 [−1.91, −1.52] | 0.00 [−0.03, 0.03] |
| Item 13 | Practical | 70.5 | 1.21 [1.06, 1.36] | −0.91 [−1.09, −0.73] | 0.01 [−0.08, 0.09] |
| Item 14 | Practical | 56.9 | 0.73 [0.63, 0.83] | −0.42 [−0.59, −0.25] | 0.00 [−0.04, 0.05] |
| Item 15 | Human | 60.5 | 1.00 [0.89, 1.12] | −0.51 [−0.60, −0.41] | 0.00 [−0.01, 0.01] |
| Item 16 | Human | 65.2 | 2.37 [1.78, 2.96] | 0.06 [−0.11, 0.23] | 0.33 [0.26, 0.40] |
| Item 17 | Human | 53.3 | 2.30 [1.77, 2.84] | 0.43 [0.30, 0.55] | 0.26 [0.21, 0.31] |
| Item 18 | Human | 33.4 | 1.10 [0.82, 1.39] | 1.10 [0.91, 1.28] | 0.08 [0.01, 0.15] |
| Item 20 | Human | 42.9 | 0.93 [0.49, 1.37] | 1.36 [1.02, 1.70] | 0.23 [0.12, 0.35] |
| Item 21 | Human | 46.8 | 1.51 [1.10, 1.91] | 0.86 [0.69, 1.03] | 0.26 [0.20, 0.32] |

*Note.* % correct is the percentage of participants who answered the item correctly. a = discrimination, b = difficulty, and g = pseudo-guessing/lower-asymptote parameter from the 3PL IRT model. n = 3,695. "Domain" indicates the intended content domain in the GenAIT blueprint. Confidence intervals (CI) were calculated using Wald's method; therefore, it is not necessarily problematic for the g parameter's lower CI to extend below zero.

IRT marginal reliability was .72, and KR-20 was .69, 95% CI [.68, .71], placing these estimates around the .70 benchmark specified for group-level research. The test information curve (Figure 3) indicated that measurement precision was greatest above the mean of the latent trait. Precision was lower at the lower end of the trait distribution. Conditional reliability increased from .55 at the 5th percentile ($\theta = -1.31$) to .63 at the 25th percentile ($\theta = -0.63$), .74 at the median ($\theta = -0.04$), .80 at the 75th percentile ($\theta = 0.58$), and .89 at the 95th percentile ($\theta = 1.52$). Conditional reliability exceeded the prespecified .70 benchmark between $\theta = -0.27$ and $\theta = 2.62$. Approximately 61% of students had latent trait estimates within this interval. These findings indicate that the scale measured average and higher levels of the trait more precisely than lower levels.

In a post hoc sensitivity analysis excluding Item 12, which showed persistent item misfit and a downturn in correct responses at higher rest scores, overall IRT model fit and associations with external variables were substantively unchanged. KR-20 decreased from .69 to .67, and marginal reliability decreased from .72 to .70 (Supplementary material S3.1.4).

We also compared GenAIT raw scores with latent trait estimates from the final 3PL model. The two scoring approaches were very strongly associated, Spearman's $\rho = .97$, 95% CI [.97, .97], $p < .001$, indicating a highly similar rank ordering of respondents. Given the greater interpretability of raw scores, they were used for descriptive reporting and subsequent correlational analyses. Sensitivity analysis revealed no substantive differences in the associations produced by raw scores and latent trait estimates (Supplementary material S3.1.4).

**Figure 3**

*GenAIT test information and latent trait distribution in the validation subsample*

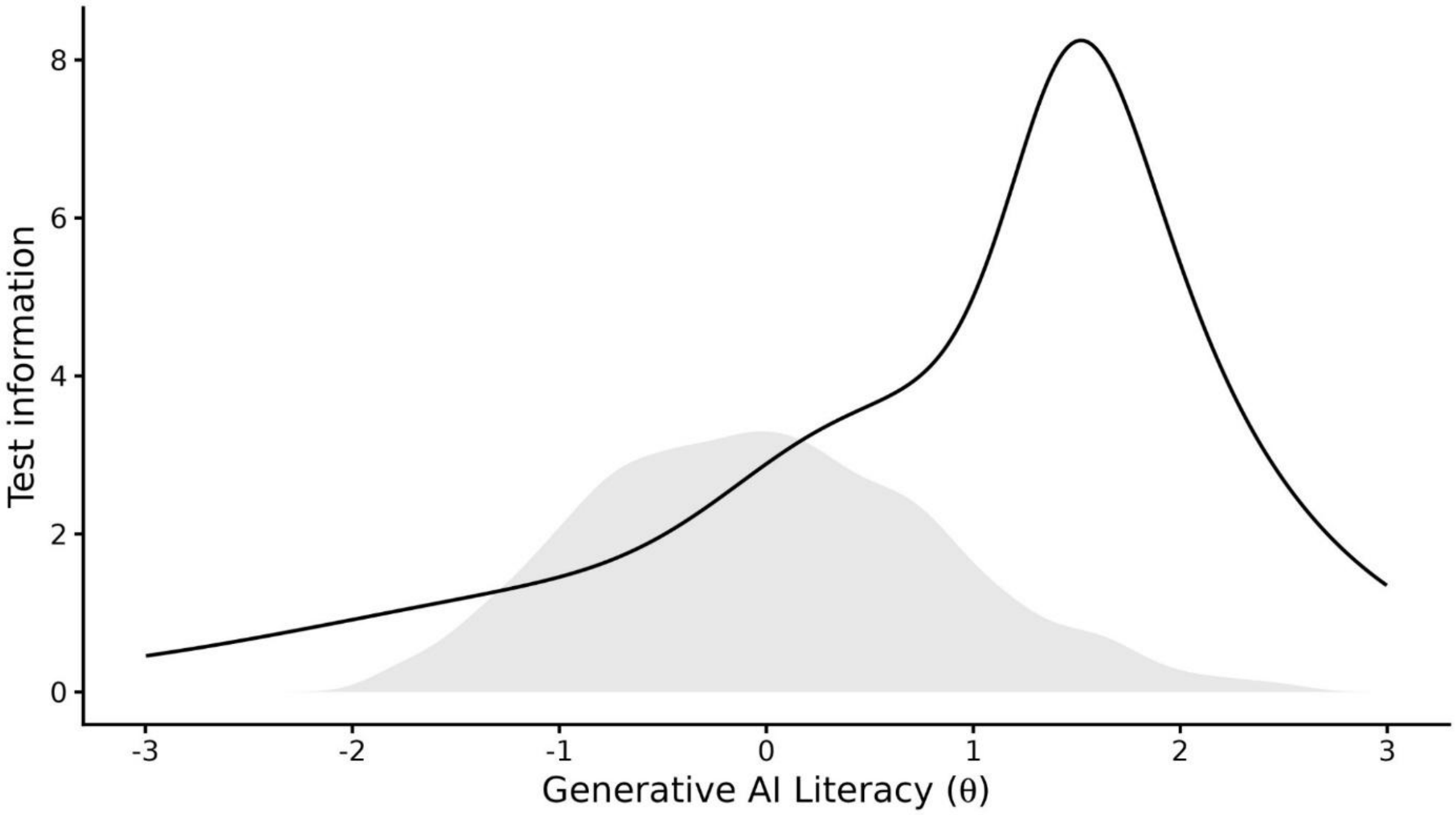

*Note.* The solid line represents the test information function for the final 18-item 3PL model. The shaded area represents the density distribution of latent trait estimates in the validation subsample. The density was rescaled for display on the test-information axis and should therefore be interpreted in terms of its shape and location rather than its height. Higher test information indicates greater measurement precision at the corresponding level of GenAI literacy.

## 5.3 Associations with external variables

After adjusting for multiple testing, GenAIT scores showed a very small negative association with perceived usefulness ($\rho = -.05$), whereas the association with perceived ease of use was not statistically significant ($\rho = .01$). GenAIT scores were also negatively associated with general ($\rho = -.16$) and school-related ($\rho = -.19$) LLM use frequency, with small to medium effect sizes (Table 4).

**Table 4**

*Spearman correlations among the study variables in the validation subsample*

| Measure | *M* | *SD* | 1 | 2 | 3 | 4 | 5 |
|---|---|---|---|---|---|---|---|
| 1. GenAIT | 8.35 | 3.39 | — | | | | |
| 2. General LLM use frequency | 4.42 | 1.54 | −.16*** [−.19, −.13] | — | | | |
| 3. School-related LLM use frequency | 4.10 | 1.45 | −.19*** [−.22, −.16] | .75*** [.74, .77] | — | | |
| 4. Perceived usefulness | 12.19 | 3.54 | −.05** [−.08, −.01] | .53*** [.50, .56] | .51*** [.48, .53] | — | |
| 5. Perceived ease of use | 13.43 | 3.00 | .01 [−.02, .04] | .36*** [.33, .39] | .33*** [.30, .36] | .62*** [.59, .64] | — |

*Note.* Spearman correlations were obtained for pairwise complete observations. Pairwise sample sizes ranged from 3,293 to 3,695. *p*-values were adjusted for multiple comparisons using the Benjamini–

Hochberg false discovery rate procedure. Three asterisks (***) indicate adjusted significance at $p < .001$, two asterisks at $p < .01$, one asterisk at $p < .05$. Brackets indicate pointwise 95% percentile bootstrap confidence intervals based on 5,000 respondent-level resamples. GenAIT included 18 multiple-choice items, and raw total scores were used for these correlations. Measures 2 and 3 included one item and were on a seven-point self-report scale. Measures 4 and 5 were a composite of three items each and measured on a six-point self-report scale (possible scores ranging from 3-18). Internal consistency for multi-item constructs was: GenAIT (marginal reliability = .72, KR-20 = .69), perceived usefulness (Cronbach's alpha = .76), and perceived ease of use (Cronbach's alpha = .69).

Sensitivity analyses comparing our original exclusion criteria with more permissive and strict criteria revealed no substantive differences in IRT model fit, associations with external variables, and score reliability (Supplementary material S3.1.5).

# 6 Discussion

## 6.1 GenAIT validity and reliability

The primary aim of this study was to develop and gather initial evidence of validity for GenAIT, an objective test intended to measure the conceptual knowledge component of GenAI literacy among high school students. Our findings provide evidence that GenAIT is a psychometrically promising instrument for group-level assessment. Below, we discuss the evidence supporting its validity through test content, internal structure, and relations with external variables, as well as evidence concerning score reliability.

To investigate RQ1, we conducted an expert review, which offered support for the content validity of GenAIT. The quantitative ratings of item relevance and clarity broadly supported the interpretation that GenAIT items are well aligned with our proposed conceptualization of GenAI literacy for high school students. Although some items underwent revisions after the review process, the changes were minor and primarily designed to improve item clarity. Evidence regarding construct coverage is more nuanced. Experts generally viewed the test as adequately covering the most important aspects of GenAI literacy, but qualitative feedback identified statistical modeling and ethics as broad categories that might be underrepresented. This suggests that the current version captures important themes but should not be considered an exhaustive representation of the broader GenAI literacy construct.

Regarding RQ2, the results provide preliminary support for the psychometric performance of the final version of GenAIT, while also identifying a mismatch in test targeting. The one-factor CFA supported approximate unidimensionality, and no problems were found with local dependence, suggesting that most item associations were sufficiently accounted for by the general conceptual-knowledge dimension. IRT analyses also demonstrated broadly acceptable item functioning and model fit, with most items showing moderate to high discrimination, broadly monotonic patterns, and only limited misfit. Together, these results suggest that a single GenAIT total score can be a reasonable summary of students' demonstrated knowledge across technical, practical, and human domains of GenAI literacy.

Score reliability of GenAIT was broadly adequate for group-level research, with KR-20 and IRT marginal reliability yielding similar estimates. These overall indices hid important variation in measurement precision across the latent trait. The test information curve showed that GenAIT measured students at average and above-average levels with considerably higher precision than lower levels. These findings warrant caution when interpreting small score differences, particularly at lower levels of GenAI literacy, and do not support high-stakes individual classifications. Future iterations of GenAIT should add easier items to improve measurement precision at lower levels of GenAI literacy.

Finally, although not an explicit research question, we found an expected association between GenAIT and magical perception of AI in a separate pilot sample. It has been proposed that novel technologies can elicit perceptions of enchantment and magic when users do not understand the mechanisms at play (Lupetti & Murray-Rust, 2024). Our results were consistent with this theoretical expectation and previous empirical work on the subject (Li et al., 2026; Tully et al., 2025), offering preliminary evidence that GenAIT behaves in an expected way in relation to at least one external construct. However, given the small pilot sample size ($n$ = 71), the estimate should be interpreted with caution, as it is too small to provide stable correlation estimates (see Schönbrodt & Perugini, 2013).

Taken together, the evidence supports GenAIT as a promising measure for research on conceptual knowledge about GenAI among Estonian high school students. It may be useful in examining the correlates of this knowledge and describing aggregate response patterns. In educational settings, GenAIT responses may help identify concepts that need further discussion or instruction. Comparison of student groups and evaluation of instructional change require further evidence about measurement invariance and sensitivity to change.

## 6.2 GenAI literacy, LLM perceptions, and use

The secondary aim of this study was to contribute to the literature on AI literacy by examining the associations between objectively measured GenAI literacy and several variables related to the TAM (RQ3). According to this framework, perceived ease of use and perceived usefulness shape both the intention to use and actual use of technologies (Davis, 1989). Favorable perceptions of LLMs have been linked to ChatGPT use in previous empirical studies (Duong et al., 2023). Consistent with this result, perceived usefulness and perceived ease of use were positively associated with LLM use frequency in the present study. Previous research has remained inconclusive about how GenAI literacy fits into the picture. In our sample, GenAI literacy showed weak or nonsignificant associations with LLM perceptions and negative associations with both general and school-related LLM use frequency.

We found no significant association of GenAI literacy with perceived ease of use, and a very small negative association with perceived usefulness. This is, to our knowledge, the first evidence of these associations from objective assessment of AI or GenAI literacy. By contrast, previous studies using self-report measures have generally found positive associations between these variables (e.g., Faiz et al., 2026; Schiavo et al., 2024). One possible explanation is fluency bias: the less cognitive friction students experience when using AI may increase perceived competence or familiarity, which may in turn be reflected in self-reported AI literacy but not demonstrated knowledge (Alter & Oppenheimer, 2009). Conceptual knowledge, when measured objectively, may be associated with opposing processes that partly cancel each other out. Increased knowledge may help students apply GenAI more effectively, while simultaneously increasing awareness of LLM limitations and the need for additional verification. These competing processes could help explain the weak associations with perceived usefulness and ease of use, but they were not directly examined in this study.

GenAI literacy was negatively associated with general LLM use frequency in the present study with a small to medium effect size. Previous studies using objective measures of AI or GenAI literacy have produced mixed findings, including no significant association (Bewersdorff et al., 2025a), a negative association (Tully et al., 2025), or a positive association (Rismanchian et al., 2026). Rismanchian et

al.'s (2026) study is most directly comparable to ours, as they also measured conceptual knowledge about generative tools and found a positive association. While previous research has proposed that AI use might not increase knowledge about these tools (Bewersdorff et al., 2025a), our cross-sectional design limits the ability to draw causal inferences. We raise the possibility that these associations depend on the target population and educational context. This may partly explain why Rismanchian et al. (2026) observed a positive association among university students, whereas we observed a negative association among high school students.

Finally, school-related LLM use was also negatively associated with GenAI literacy, with a small to medium effect size. One possible explanation is that students with higher GenAI literacy are more aware of potential limitations and risks associated with using LLMs for learning. Evidence from AI-assisted learning suggests that if thinking is routinely delegated to LLMs through cognitive offloading (see Risko & Gilbert, 2016), later unaided performance may suffer (Bastani et al., 2025). Students with higher GenAI literacy may therefore be more selective about using LLMs for schoolwork. Nevertheless, this interpretation remains speculative, as selectivity cannot be inferred from use frequency alone.

## 6.3 Educational implications

Findings from our main validation study suggest several educational implications. Most notably, students could benefit from additional instruction, learning opportunities, or educational support to develop a stronger foundational understanding of what GenAI is, how it works, and how it may impact individuals and society. Additionally, frequent GenAI use and favorable perceptions should not be assumed to indicate conceptual understanding of GenAI.

Students answered many items incorrectly, particularly those in the technical content domain. This suggests that many students may hold inaccurate or incomplete mental models of the concepts sampled by GenAIT. This is broadly consistent with previous observations about university students often anthropomorphizing AI or treating it as a search engine (Rismanchian et al., 2026). At the same time, students reported frequent LLM use, with the average student using these tools several times a week, including for schoolwork. This combination of frequent use and difficulty answering many conceptual knowledge items is particularly important as students' GenAI-related learning practices and habits are still emerging, and pedagogical responses remain under development (Bardone & Forsler, 2026). Limited understanding may contribute to misplaced trust or over-reliance on AI systems (Clerc et al., 2026), which may in turn undermine later unaided learning or performance (Bastani et al., 2025). Consistent with these concerns, previous experimental work has reported substantial over-reliance among Estonian high school students (Puppart & Aru, 2025).

Collectively, these findings highlight educational needs that programs such as AI Leap (AI Leap Foundation, 2025) could address to help students benefit from GenAI. Importantly, such initiatives should go beyond teaching students how to operate current tools or formulate effective prompts and address more foundational concepts, such as how GenAI produces outputs, why critical verification is important, and what ethical and societal considerations are related to GenAI.

# 7 Limitations

Several limitations must be considered when interpreting the current findings. First, GenAIT is designed to measure the conceptual knowledge component of GenAI literacy. Therefore, the results presented here might not generalize to all components of GenAI literacy, including skills and the application of knowledge in authentic tasks.

Second, some psychometric limitations remain regarding GenAIT. The test did not measure all students with equal precision, and some items were flagged for low factor loadings or discrimination, item misfit, or possible deviations from monotonicity. While we kept some of the flagged items considering the magnitude of the deviation and content validity, they should be investigated and improved in future iterations of GenAIT. Additionally, easier items should be added to improve precision at lower levels of the trait.

Third, evidence based on associations with external variables remains limited. The primary evidence for theoretically expected associations came from a pilot study with a small and potentially selective sample. Further research is therefore needed to establish convergent, discriminant, criterion-related, and predictive validity.

Fourth, we could not examine differential item functioning or broader measurement invariance across grade levels or gender due to data-protection constraints. Although the 3PL IRT model demonstrated adequate overall fit and no problematic local dependence, generalization across student subgroups cannot yet be assumed and must be examined in future studies. Because school identifiers were unavailable, we also could not account for clustering of students within schools. Consequently, the reported standard errors, confidence intervals, and significance tests do not account for possible dependence among students attending the same school.

Fifth, further work is needed to strengthen evidence based on test content and response processes. The items were designed to reflect plausible misconceptions identified through literature review and researcher judgment, but students' interpretations were not directly examined. Cognitive interviews or think-aloud protocols could help establish whether items and distractors are understood as intended. Content coverage should also be revisited because the final test samples only part of a broad and continuously evolving construct. Experts identified some potentially missing concepts, and future versions should therefore undergo renewed expert review, with items periodically revised or replaced to keep pace with developments in the field.

Finally, the associations between GenAIT, LLM perceptions, and LLM use frequency were correlational. Therefore, our ability to draw causal inferences is limited. Future research should use experimental and longitudinal designs to clarify the direction and mechanisms underlying these associations.

## 8 Conclusion

This study developed GenAIT, a multiple-choice test measuring the conceptual knowledge component of GenAI literacy, and provided initial validity evidence in a large sample of Estonian high school students. Evidence from test content, internal structure, item functioning, and score reliability supports its use for group research, although lower precision at lower trait scores limits interpretation and motivates the addition of easier items. GenAIT was not significantly associated with perceived ease of use, negatively associated with perceived usefulness with a very small effect size, and negatively associated with both general and school-related LLM use with a small to medium effect size. These results highlight the importance of distinguishing AI adoption and favorable attitudes from demonstrated conceptual understanding of generative technologies.

## 10 Funding

This work was supported by the Estonian Research Council grants PSG728 and Tem-TA 120 and the Estonian Centre of Excellence in Artificial Intelligence (EXAI), funded by the Estonian Ministry of Education and Research.

## 11 Acknowledgments

We are grateful to all the experts who participated in the content validation process, as well as Madis Vasser, Kaidi Tootmaa, and Mario Mäeots whose comments and ideas during item generation and refinement were very helpful.